\documentclass[12pt,a4paper,final]{iopart}
\expandafter\let\csname equation*\endcsname\relax
\expandafter\let\csname endequation*\endcsname\relax
\usepackage{iopams}
\usepackage{graphicx}
\usepackage{subfigure}
\usepackage[breaklinks=true,colorlinks=true,linkcolor=blue,urlcolor=blue,citecolor=blue]{hyperref}
\usepackage{cite}
\usepackage{braket}
\usepackage{mathtools}

\usepackage[a4paper]{geometry}
\usepackage{bpchem,upgreek}
	\usepackage{amsmath}
	\usepackage{makeidx}
	\usepackage{amsfonts}
	\usepackage[ansinew]{inputenc}
	\usepackage[usenames,dvipsnames]{pstricks}
    \usepackage{graphicx}
    \usepackage{float}
    \usepackage{subfigure}
	\usepackage{pst-grad} 
	\usepackage{pst-plot} 
	\usepackage{sidecap}
	\usepackage[makeroom]{cancel}

\begin{document}

\title[Quantum repeater protocol...]{Quantum repeater protocol using an arrangement  of QED-optomechanical hybrid systems}

\author{M Ghasemi$^{1}$}
\author{M K Tavassoly$^{1}$}
\address{$^1$Atomic and Molecular Group, Faculty of Physics, Yazd University, Yazd  89195-741, Iran}
\ead{mktavassoly@yazd.ac.ir}

\vspace{10pt}
\date{today}

\begin{abstract}
In this paper we consider the quantum repeater protocol for distributing the entanglement to two distant three-level atoms. In this protocol, we insert six atoms between two target atoms such that the eight considered atoms are labeled by $1,2,\cdots 8$, while only each two adjacent atoms $(i,i+1)$ with $i=1,3,5,7$ are entangled. Initially, the separable atomic pair states (1,4) and (5,8) become entangled by performing interaction between atoms (2,3) and (6,7) in two optomechanical cavities, respectively. Then, via performing appropriate interaction between atoms (4,5) in an optical cavity quantum electrodynamics (QED) approach, the target atoms (1,8) are finally become entangled. Throughout this investigation, the effects of mechanical frequency and optomechanical coupling strength to the field modes on the produced entanglement and the related success probability are evaluated. It is shown that, the time period of produced entanglement can be  developed by increasing the mechanical frequency. Also, maximum of success probability of atoms (1,8) is increased by decreasing the optomechanical coupling strength to the field modes in most cases.
\end{abstract}


\vspace{2pc}
%
%
%
%

\section{Introduction}
It is recently approved that distributing entanglement and entangled states to long distances is possible by using quantum repeater protocol \cite{Briegel1998,Yi2019,Uphoff2016,Bernad2013,Ghasemi20193}. A scheme for quantum repeaters related to pure two-mode squeezed states over long distances has been presented in \cite{Furrer2018}. Heralded quantum repeater based on quantum dots has been investigated in \cite{Li2016}. Recently, we have considered the quantum repeater based on atoms where the entanglement has been swapped between atoms via performing interaction in the optical cavities \cite{,Ghasemi2018,Ghasemi2019,Ghasemi20192}. Quantum repeater is very useful in quantum internet issue \cite{Azuma2017}. In  \cite{Behera2019} the entanglement purification and swapping protocol have been considered to design quantum repeater in quantum computer. In this protocol a long distance is divided to several short parts and entanglement is transferred by entanglement swapping process. The entanglement can be swapped by using Bell state measurement (BSM) \cite{Ghasemi2016,Ghasemi2017,Nourmandipour2016,Liao2011} and also via performing interaction between separable parts of the considered system \cite{Pakniat2017,Soltani2017}. The quantum electrodynamics (QED) method is indeed more complicated than the BSM, at least from the theoretical point of view. However, the realization and discernment of Bell states which is necessary in the BSM method is still practically difficult \cite{Xue2006,Yang2005,Yang2006,Osnaghi2001}, even though this method is usually in the core of entanglement swapping processes in the (theoretical) literature deal with entanglement swapping. In relation to the mentioned fact, it may be recalled that, entanglement swapping without BSM method is also of enough interest for people who work in this field \cite{Cardoso2009,Giang2010,Xiu2007,Pakniat2017}. The quantum interactions can be modeled by Jaynes-Cummings model (JCM) \cite{Jaynes1963,Bashkirov2006,Gerry1992,Bartzis1991,Obada2004,Hu1989} as well as Tavis-Cummings model (TCM) \cite{Tavis1968,Bashkirov2012}. Based on these models, the interactions can be performed in optomechanical cavity \cite{Nadiki2016,Nadiki2018}, instead of the usually used optical cavity.\\ Nowadays, cavity optomechanics is a noticeable field of physics with its relevant characteristics, namely, the interaction between a mechanical resonator and radiation pressure of an optical cavity field  which attracted a lot of attention \cite{Caves1980,Corbitt2006}. In this way, many theoretical \cite{Huang2010,Joshi2012,Barzanjeh2011} as well as experimental \cite{Safavi2011,Verhagen2012,Anetsberger2009} analyzes have been performed in this line of research. Recently, some particular features of such systems have been discussed  in \cite{Zhang2017,Eghbali2017,Feng2016}. The possible potential application of optomechanical system has been proposed in \cite{Galland2014}. In Ref. \cite{Vitali2007}, the authors have shown that, how an optical cavity field mode become entangled with a macroscopic vibrating mirror by means of radiation pressure.
 In \cite{Aspelmeyer2014} the authors have presented the quantum theory of optomechanical cooling and outlined the possible applications of optomechanical devices in laser sciences. The simplest arrangement of an optomechanical system includes of a Fabry-Perot cavity in which one mirror of the cavity is movable under the influence of radiation pressure of the light inside the cavity \cite{Aspelmeyer2014}. In fact, when a resonant optical laser field is sent into such an optomechanical cavity, as is well-known the radiation exerts a force on the mechanical resonator and causes it to oscillate. These mechanical oscillations modify the length of optomechanical cavity and thus its resonant frequency. Consequently, a coupling between the optical and mechanical degrees of freedom  is created
 \cite{Wang2014}. Accordingly, the basic problem in such systems is to control the motion of oscillating mirror with the light inside the cavity. This idea has been used to a broad range of scales from macroscopic mirrors in \cite{Corbitt2007} to nano- or micro-mechanical cantilevers \cite{Metzger2004,Kleckner2006}, micro-toroids \cite{Carmon2005}, membranes \cite{Thompson2008} and optomechanical crystals \cite{Eichenfield2009}.
 As stated in \cite{Aspelmeyer2012}, cavity optomechanical devices range from nanometer-sized devices of as little as $10^7$ atoms to micromechanical
 structures of $10^{14}$ atoms and to macroscopic centimeter-sized mirrors. The authors also showed that, optomechanical systems can be helpful to measure feeble forces and fields with a sensitivity, precision and accuracy approaching the quantum limit.
 As a result, the optomechanical systems can provide macroscopic systems by which one can find some quantum mechanical phenomena inherent in their structural mechanisms. 
 Quantum repeater protocols have been considered in the presence of optical cavities in the previous literature \cite{Yi2019,Ghasemi2018,Ghasemi2019,Ghasemi20192,van2008,Wang2012,Ladd2006}, moreover, the main feature of our present work is to extend, even partially, a particular quantum information protocol, in this case, the quantum repeaters, to the macroscopic quantum systems instead of microscopic quantum systems.\\
In the present paper we want to consider the quantum repeater based on three-level V-type atomic states, in which entanglement can be transferred to two distant atoms. In this protocol by inserting six atoms between the two target atoms, the long distance is divided to four short parts. Therefore, this investigation for eight three-level atoms $(1,2,\cdots 8)$ is considered in three statuses. Initially, the pairs $(i,i+1)$ where $i=1,3,5,7$, are prepared in appropriate entangled states; then, by performing interaction between atoms (2,3) and (6,7) in separate optomechanical cavities (see figure \ref{fig.Fig1a}), entanglement is respectively generated between atoms (1,4) and (5,8). Finally, the entangled state of target atoms (1,8) can be achieved by performing appropriate interaction, however, in an optical cavity (see figure \ref{fig.Fig1a}). Henceforth, the noticeable feature of our work is considering QED-optomechanics hybrid system in the presence of eight three-level atoms to construct the quantum repeater protocol.
  We are trying to fully extend the optomechanical systems instead of optical cavities in a quantum repeater which the results can be published elsewhere.\\
 This paper is organized as follows: Our quantum repeater protocol is introduced in Sec. 2. The numerical results for entropy and success probability are presented in Sec. 3. At last,
  summary and conclusions of the paper are gathered in Sec. 4.

\section{Quantum repeater protocol}\label{model}
 \begin{figure}[H]
  \centering
     \subfigure[\label{fig.Fig1a} \ The quantum repeater protocol]{\includegraphics[width=0.44\textwidth]{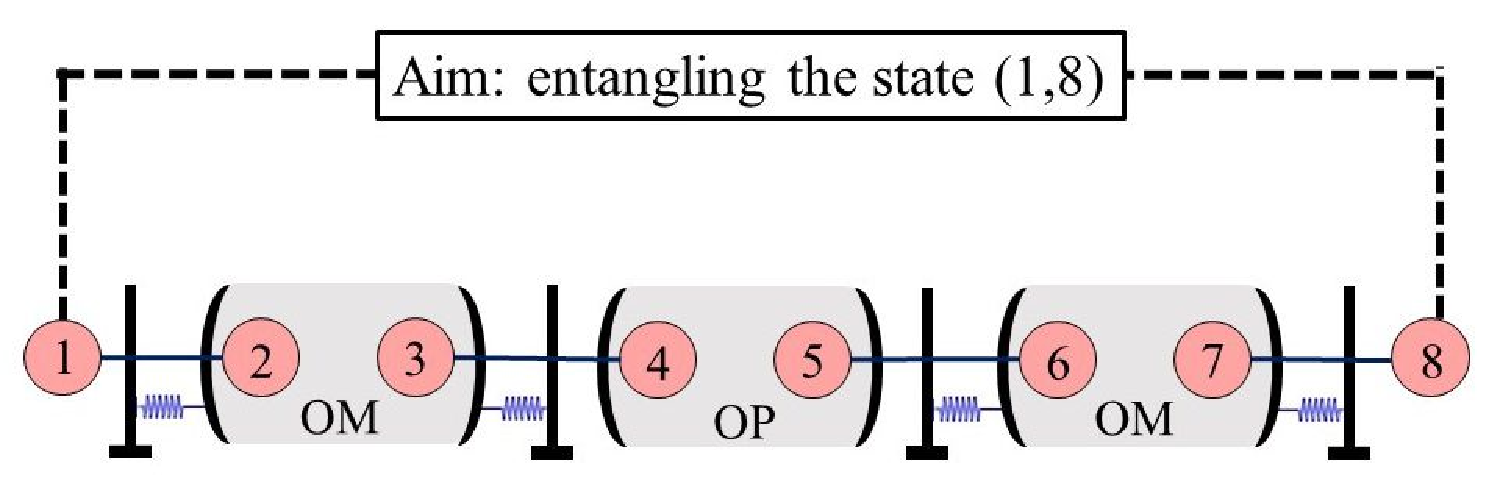}}
     \hspace{0.05\textwidth}
     \subfigure[\label{fig.Fig1b} \ The three-level V-type atoms]{\includegraphics[width=0.22\textwidth]{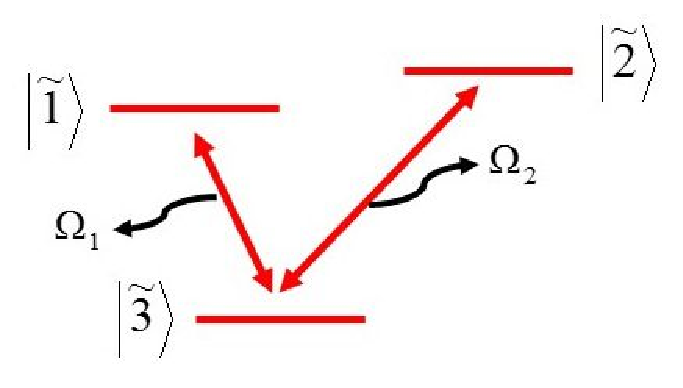}}
\hspace{0.05\textwidth}
     \subfigure[\label{fig.Fig1c} \ Sketch of the OM system, where $j=2,6$]{\includegraphics[width=0.22\textwidth]{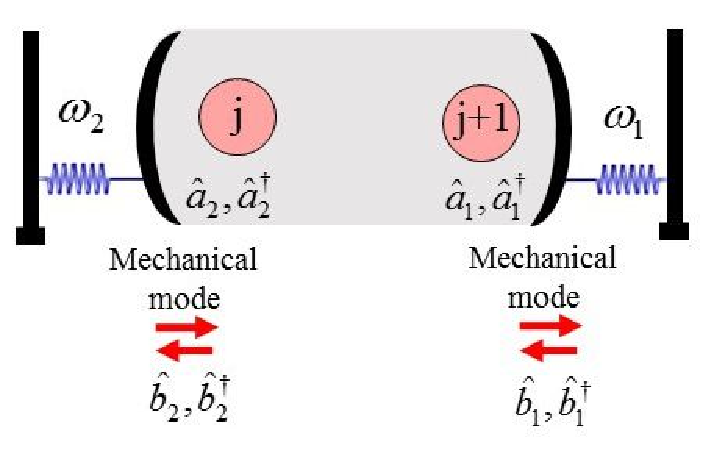}}
   \caption{\label{fig:Fig1} {(a) The used scheme of quantum repeater protocol. OM and OP refer to optomechanical system and optical cavity system, respectively. (b) The diagram of three-level V-type atoms $(1,2,\cdots 8)$. (c) Sketch of the OM cavity with two movable mirrors.}}
  \end{figure}

 As shown in figure \ref{fig.Fig1a}, the aim of our consideration is distributing the entanglement to two distant separable atoms (1,8). By using quantum repeater protocol, the target pair (1,8) is converted to entangled state in three statuses. Initially, only the adjacent pairs of $(i,i+1)$ where $i=1,3,5,7$, have been prepared in appropriate entangled states, then the separable atoms (1,4) and (5,8) are entangled by performing interaction between atoms (2,3) as well as (6,7) in two optomechanical systems. Finally, our purpose, \textit{i.e.}, distributing the entanglement to atoms (1,8), is achieved by performing interaction of the atoms (4,5) in an optical cavity.\\
  We first consider the V-type atoms (1,2,3,4), with energy levels shown in figure \ref{fig.Fig1a}, in such a way that the pairs of (1,2) and (3,4) have been prepared in the following entangled state (see figure \ref{fig.Fig1b}):
   \begin{eqnarray}\label{initialstate}
\ket{\psi(0)}_{i,i+1}&=&\frac{1}{\sqrt{2}}(\ket{\widetilde{1},\widetilde{3}}+\ket{\widetilde{3},\widetilde{1}})_{i,i+1}, \qquad   i=1,3.
  \end{eqnarray}
 Consider two V-type atoms labeled with (2,3) with atomic transition operator $\hat{\sigma}_{lm}^k=\ket{\widetilde{l}}_k\bra{\widetilde{m}}$ where $l,m=1,2,3$ denote the three levels of atoms, $k=2,3$ refer to the atomic numbers with atomic frequencies $\widetilde{\omega}_i,\widetilde{\omega}^{'}_i$, respectively in the presence of two photon modes $\hat{a}_j$, $\hat{a}^{\dagger}_j$ and two phonon modes $\hat{b}_j$, $\hat{b}^{\dagger}_j$, where $j=1,2$ (see figure \ref{fig.Fig1c}). Frequencies of photon (phonon) modes are denoted as $\Omega_1$, $\Omega_2$ ($\omega_1$, $\omega_2$).
 Based on the described model we perform the three partite atom-optics-optomechanics interaction with the  
  Hamiltonian $\hat{H}_{(2,3)}=\hat{H}_0+\hat{H}_1$ where $(\hbar=1)$
\begin{eqnarray}\label{hamiltonian}
 \hat{H}_0&=&\sum^3_{i=1}\widetilde{\omega}_i\hat{\sigma}_{ii}^2+\sum^3_{i=1}\widetilde{\omega}^{'}_i\hat{\sigma}_{ii}^3+\sum_{j=1,2}\Omega_j\hat{a}_j^{\dagger}\hat{a}_j+\sum_{j=1,2}\omega_j\hat{b}_j^{\dagger}\hat{b}_j,\\\nonumber
 \hat{H}_1&=&\lambda_1(\hat{a}_1\hat{\sigma}_{13}^2+\hat{a}_1^{\dagger}\hat{\sigma}_{31}^2)+\lambda_2(\hat{a}_2\hat{\sigma}_{23}^2+\hat{a}_2^{\dagger}\hat{\sigma}_{32}^2)+\lambda^{'}_1(\hat{a}_1\hat{\sigma}_{13}^3+\hat{a}_1^{\dagger}\hat{\sigma}_{31}^3)\\\nonumber
 &+&\lambda^{'}_2(\hat{a}_2\hat{\sigma}_{23}^3+\hat{a}_2^{\dagger}\hat{\sigma}_{32}^3)-G\hat{a}_1^{\dagger}\hat{a}_1(\hat{b}_1+\hat{b}_1^{\dagger})-G^{'}\hat{a}_2^{\dagger}\hat{a}_2(\hat{b}_2+\hat{b}_2^{\dagger}).
  \end{eqnarray}
In Hamiltonian (\ref{hamiltonian}), $\lambda_i$, $\lambda^{'}_i$ ($G$, $G^{'}$) are the coupling constants between atoms (mechanical modes) and optical modes. The phonon emission experiments measure directly the frequency and angular distribution for generating different phonon modes \cite{Zhou2011,Xu1996}. In \cite{Nishioka2014}, two-phonon modes were strongly coupled by external fields via two-phonon excitation. In this study differential frequency between two external fields have been tuned to differential frequency of the two phonon modes. The Hamiltonian (\ref{hamiltonian}) in the interaction picture may be obtained as:
   \begin{eqnarray}\label{hbl}
   \hat{H}^\mathrm{int}_{(2,3)}&=&e^{i \hat{H}_0 t} \hat{H}_1e^{-i \hat{H}_0 t}\\ \nonumber
   &=&\hat{H}_1+it\left[\hat{H}_0,\hat{H}_1 \right]+\frac{(it)^2}{2!} \left[\hat{H}_0,\left[\hat{H}_0,\hat{H}_1 \right]\right]+\cdots\\\nonumber
   &=&e^{i\omega_M t}\left[ \lambda_1\left( \hat{a}_1\hat{\sigma}^2_{13}+\hat{a}_1\hat{\sigma}^3_{13}\right)+\lambda_2\left( \hat{a}_2\hat{\sigma}^2_{23}+\hat{a}_2\hat{\sigma}^3_{23}\right)\right. \\\nonumber
   &-&G\left( \left.  \hat{a}_1^{\dagger}\hat{a}_1\hat{b}^{\dagger}_1+ \hat{a}_2^{\dagger}\hat{a}_2\hat{b}^{\dagger}_2\right) \right] + h.c.,
       \end{eqnarray}
where we assumed\footnote{The procedure can push forward without these assumptions, in fact, for simplification of equations we applied these assumptions.} that $\lambda_1=\lambda^{'}_1$, $\lambda_2=\lambda^{'}_2$, $G=G^{'}$, $\widetilde{\omega}_i=\widetilde{\omega}^{'}_i$, $\omega_1=\omega_2=\omega_M$ and $\widetilde{\omega}_1-\widetilde{\omega}_3-\Omega_1=\omega_M=\widetilde{\omega}_2-\widetilde{\omega}_3-\Omega_2$ (which needs, at first, to tune the frequency of the fields such that $\widetilde{\omega}_1-\Omega_1=\widetilde{\omega}_2-\Omega_2$, after choosing the three-level atom).
The effective Hamiltonian, then can be achieved by using (\ref{hbl}) and following the path of Refs. \cite{James2007,Gamel2010}, results in:
\begin{eqnarray}\label{effectivehamiltonian}
\hat{H}^{\mathrm{eff}}_{(2,3)}&=&\frac{\lambda^2_1}{\omega_M}\sum_{i=2,3}\left[ \hat{\sigma}_{11}^i+\hat{a}_1^{\dagger}\hat{a}_1\left( \hat{\sigma}_{11}^i-\hat{\sigma}_{33}^i\right) \right]\\\nonumber
&+&\frac{\lambda^2_2}{\omega_M}\sum_{i=2,3}\left[ \hat{\sigma}_{22}^i+\hat{a}_2^{\dagger}\hat{a}_2\left( \hat{\sigma}_{22}^i-\hat{\sigma}_{33}^i\right) \right] \\\nonumber
&+&\frac{\lambda^2_1}{\omega_M}\left( \hat{\sigma}_{13}^2 \hat{\sigma}_{31}^3+\hat{\sigma}_{31}^2 \hat{\sigma}_{13}^3\right) +\frac{\lambda^2_2}{\omega_M}\left( \hat{\sigma}_{23}^2 \hat{\sigma}_{32}^3+\hat{\sigma}_{32}^2 \hat{\sigma}_{23}^3\right)\\\nonumber
&-&\frac{G^2}{\omega_M}\left[\left(\hat{a}_1^{\dagger}\hat{a}_1 \right)^2 +\left(\hat{a}_2^{\dagger}\hat{a}_2 \right)^2 \right]+\frac{\lambda_1\lambda_2}{\omega_M}\left( \sum_{i=2,3}\hat{a}_1\hat{a}^{\dagger}_2  \hat{\sigma}_{12}^i+h.c.\right) \\\nonumber
& -&\frac{G \lambda_1}{\omega_M}\left( \sum_{i=2,3}\hat{a}_1\hat{b}_1  \hat{\sigma}_{13}^i+h.c.\right)-\frac{G \lambda_2}{\omega_M}\left( \sum_{i=2,3}\hat{a}_2\hat{b}_2  \hat{\sigma}_{23}^i+h.c.\right).
\end{eqnarray}
 Keeping in mind Eq. (\ref{initialstate}), the initial states of atoms (1,2,3,4) and modes $(a_1,a_2;b_1,b_2)$ are respectively denoted by $\ket{\psi(0)}_{1,2}\otimes\ket{\psi(0)}_{3,4}$ and $\ket{0,0;0,0}$, \textit{i.e.,} we assumed the optical and mechanical modes are all in vacuum states. Then, with the help of effective Hamiltonian (\ref{effectivehamiltonian}) and the time-dependent Schr$\mathrm{\ddot{o}}$dinger equation $i\frac{\partial}{\partial t}\ket{\psi(t)}=\hat{H}^{\mathrm{eff}}_{(2,3)}\ket{\psi(t)}$, the entangled state related to atoms (1,2,3,4) and field modes can be stated as below,
\begin{eqnarray}\label{state1-4}
 &&\ket{\psi(t)}\\\nonumber
 &=&A_1(t)\ket{0,0;0,0;\widetilde{1},\widetilde{3};\widetilde{3},\widetilde{1}}+A_2(t)\ket{0,0;0,0;\widetilde{1},\widetilde{3};\widetilde{1},\widetilde{3}}\\\nonumber
  &+&A_3(t)\ket{0,0;0,0;\widetilde{1},\widetilde{1};\widetilde{3},\widetilde{3}}+A_4(t)\ket{1,0;1,0;\widetilde{1},\widetilde{3};\widetilde{3},\widetilde{3}}\\\nonumber
   &+&A_5(t)\ket{0,0;0,0;\widetilde{3},\widetilde{1};\widetilde{1},\widetilde{3}}+A_6(t)\ket{1,0;1,0;\widetilde{3},\widetilde{3};\widetilde{1},\widetilde{3}}\\\nonumber
  &+&A_7(t)\ket{1,0;1,0;\widetilde{3},\widetilde{1};\widetilde{3},\widetilde{3}}+A_8(t)\ket{2,0;2,0;\widetilde{3},\widetilde{3};\widetilde{3},\widetilde{3}}\\\nonumber
   &+&A_9(t)\ket{0,0;0,0;\widetilde{3},\widetilde{1};\widetilde{3},\widetilde{1}}+A_{10}(t)\ket{0,0;0,0;\widetilde{3},\widetilde{3};\widetilde{1},\widetilde{1}}\\\nonumber
    &+&A_{11}(t)\ket{1,0;1,0;\widetilde{3},\widetilde{3};\widetilde{3},\widetilde{1}}.
\end{eqnarray}
In the continuation, we calculate the coefficients introduced in (\ref{state1-4}). Utilizing the effective Hamiltonian (\ref{effectivehamiltonian}) and the proposed state (\ref{state1-4}) in the time-dependent Schr$\mathrm{\ddot{o}}$dinger equation, the uncoupled equation $\dot{A}_1(t)=0$ and the following three sets of coupled differential equations
 \begin{eqnarray}\label{diffeq2}
 \dot{A}_2(t)&=&-i\frac{\lambda^2_1}{\omega_M}\left( A_2(t)+A_3(t)\right)+i\frac{G\lambda_1}{\omega_M}A_4(t),\\\nonumber
 \dot{A}_3(t)&=&-i\frac{\lambda^2_1}{\omega_M}\left( A_2(t)+A_3(t)\right)+i\frac{G\lambda_1}{\omega_M}A_4(t),\\\nonumber
 \dot{A}_4(t)&=&i\frac{G\lambda_1}{\omega_M}\left( A_2(t)+A_3(t)\right)+\frac{i}{\omega_M}\left( 2\lambda^2_1+G^2\right) A_4(t),
  \end{eqnarray}
  \begin{eqnarray}\label{diffeq3}
  \dot{A}_5(t)&=&-2i\frac{\lambda^2_1}{\omega_M}A_5(t)+i\frac{G\lambda_1}{\omega_M}\left( A_6(t)+A_7(t)\right) ,\\\nonumber
  \dot{A}_6(t)&=&i\frac{G\lambda_1}{\omega_M} A_5(t)-\frac{i}{\omega_M}\left( \lambda^2_1-G^2\right)A_6(t)\\\nonumber
  &-&i\frac{\lambda^2_1}{\omega_M}A_7(t)+2i\frac{G\lambda_1}{\omega_M}A_8(t),\\\nonumber
   \dot{A}_7(t)&=&i\frac{G\lambda_1}{\omega_M} A_5(t)-i\frac{\lambda^2_1}{\omega_M}A_6(t)\\\nonumber
     &-&\frac{i}{\omega_M}\left( \lambda^2_1-G^2\right)A_7(t)+2i\frac{G\lambda_1}{\omega_M}A_8(t),\\\nonumber
  \dot{A}_8(t)&=&2i\frac{G\lambda_1}{\omega_M}\left( A_6(t)+A_7(t)\right)+\frac{4i}{\omega_M}\left( \lambda^2_1+G^2\right) A_8(t),
   \end{eqnarray}
   and
   \begin{eqnarray}\label{diffeq4}
   \dot{A}_9(t)&=&-i\frac{\lambda^2_1}{\omega_M}(A_9(t)+A_{10}(t))+i\frac{G\lambda_1}{\omega_M}A_{11}(t),\\\nonumber
   \dot{A}_{10}(t)&=&-i\frac{\lambda^2_1}{\omega_M}(A_9(t)+A_{10}(t))+i\frac{G\lambda_1}{\omega_M}A_{11}(t),\\\nonumber
   \dot{A}_{11}(t)&=&i\frac{G\lambda_1}{\omega_M}\left( A_9(t)+A_{10}(t)\right)+\frac{i}{\omega_M}\left( 2\lambda^2_1+G^2\right) A_{11}(t),
    \end{eqnarray}
    can be obtained. The equation $\dot{A}_1(t)=0$ easily results in $A_1(t)=\frac{1}{2}$. Also, paying attention to the considered initial conditions as well as the equations (\ref{diffeq2}) and (\ref{diffeq4}), we find that $A_2(t)=A_9(t)$, $A_3(t)=A_{10}(t)$, $A_4(t)=A_{11}(t)$ and from equation (\ref{diffeq3}) we find that $A_6(t)=A_{7}(t)$. To calculate the coefficients $A_4(t)$, $A_5(t)$ and $A_8(t)$ one should solve the first and second sets of the above equations, \textit{i.e.,} equations in (\ref{diffeq2}), (\ref{diffeq3}). From equations in (\ref{diffeq2}) one readily obtains:
     \begin{eqnarray}\label{diffequ31}
                  \dot{\widetilde{A}}_{2,3}(t)&=&2i\frac{G\lambda_1}{\omega_M} \widetilde{A}_4(t) e^{i\frac{(4 \lambda^2_1+G^2)}{\omega_M}t},\\\nonumber
                  \dot{\widetilde{A}}_4(t)&=&i\frac{G\lambda_1}{\omega_M}\widetilde{A}_{2,3}(t)e^{-i\frac{(4 \lambda^2_1+G^2)}{\omega_M}t},
           \end{eqnarray}
      where we assumed,
       \begin{eqnarray}\label{diffequa3}
                 \widetilde{A}_{2,3}(t)&=&A_{2,3}(t)e^{i\frac{2\lambda^2_1}{\omega_M}t},\\\nonumber
          A_{2,3}(t)&=&A_2(t)+A_3(t),\\\nonumber
           \widetilde{A}_4(t)&=&A_4(t)e^{-i\frac{(2\lambda^2_1+G^2)}{\omega_M}t}.
           \end{eqnarray}
 Also, from equations in (\ref{diffeq3}) we have
     \begin{eqnarray}\label{diffequ3}
     \dot{\widetilde{A}}_5(t)&=&i\frac{G\lambda_1}{\omega_M}\widetilde{A}_{6,7}(t)e^{i\frac{G^2}{\omega_M}t},\\\nonumber
     \dot{\widetilde{A}}_{6,7}(t)&=&2i\frac{G\lambda_1}{\omega_M} \widetilde{A}_5(t)e^{-i\frac{G^2}{\omega_M}t}+4i\frac{G\lambda_1}{\omega_M}\widetilde{A}_8(t) e^{i\frac{\left( 6\lambda^2_1+3G^2\right)}{\omega_M} t},\\\nonumber
     \dot{\widetilde{A}}_8(t)&=&2i\frac{G\lambda_1}{\omega_M}\widetilde{A}_{6,7}(t) e^{-i\frac{\left( 6\lambda^2_1+3G^2\right)}{\omega_M} t},
      \end{eqnarray}
 where we assumed,
  \begin{eqnarray}\label{diffequa3}
    \widetilde{A}_5(t)&=&A_5(t)e^{2i\frac{\lambda^2_1}{\omega_M}t},\\\nonumber
     \widetilde{A}_{6,7}(t)&=&A_{6,7}(t)e^{i\frac{\left(2\lambda^2_1-G^2 \right)}{\omega_M}t},
         \\\nonumber
     A_{6,7}(t)&=&A_6(t)+A_7(t),\\\nonumber
          \widetilde{A}_8(t)&=&A_8(t)e^{-4i\frac{\left( \lambda^2_1+G^2\right)}{\omega_M}t}.
      \end{eqnarray}
 Now, using laplace transform to solve equations (\ref{diffequ31}), (\ref{diffequ3}), the coefficients $A_4(t)$, $A_5(t)$, $A_8(t)$ are calculated.
Also, the coefficients $A_2(t)$, $A_3(t)$ and $A_6(t)$ can be calculated.
Now, by applying the projection operators which are performed with the states $\ket{0;0} \otimes \ket{\widetilde{3};\widetilde{1}} $, $\ket{0;0} \otimes \ket{\widetilde{1};\widetilde{3}} $ and $\ket{1;1} \otimes \ket{\widetilde{3};\widetilde{3}} $ related to the state of field modes ($a_1;b_1$) and the state of atomic pair $(2;3)$ on state (\ref{state1-4}), the atoms (1,4) are respectively converted to the following entangled states,
\begin{eqnarray}\label{state114}
    \ket{\psi(t)}^1_{1,4}&=&\frac{1}{\sqrt{P^1_{1,4}(t)} }(A_2(t)\ket{\widetilde{1},\widetilde{3}}+A_{10}(t)\ket{\widetilde{3},\widetilde{1}}),
 \end{eqnarray}
 and
 \begin{eqnarray}\label{state214}
     \ket{\psi(t)}^2_{1,4}&=&\frac{1}{\sqrt{P^1_{1,4}(t)} }(A_{10}(t)\ket{\widetilde{1},\widetilde{3}}+A_{2}(t)\ket{\widetilde{3},\widetilde{1}}),
  \end{eqnarray}
 (notice that $A_2(t)=A_9(t)$, $A_3(t)=A_{10}(t)$) that the entropy and success probability have been obtained as
 \begin{eqnarray}\label{ent14}
     E^1_{1,4}(t)&=&1-\frac{\left| A_2(t)\right|^4+\left| A_{10}(t)\right|^4}{(P^1_{1,4}(t))^2},
   \end{eqnarray}
\begin{eqnarray}\label{suc14}
   P^1_{1,4}(t)&=&\left| A_2(t)\right|^2+\left| A_{10}(t)\right|^2,
 \end{eqnarray}
   and
 \begin{eqnarray}\label{state314}
     \ket{\psi(t)}^3_{1,4}&=&\frac{1}{\sqrt{2} }(\ket{\widetilde{1},\widetilde{3}}+\ket{\widetilde{3},\widetilde{1}}),
  \end{eqnarray}
   (notice that $A_4(t)=A_{11}(t)$) with the entropy and success probability have been obtained as
     \begin{eqnarray}
          E^2_{1,4}(t)&=&0.5,
        \end{eqnarray}
 \begin{eqnarray}\label{suc214}
    P^2_{1,4}(t)&=&2\left| A_4(t)\right|^2.
  \end{eqnarray}
%
%
The above-mentioned process can be repeated for atoms (5,6,7,8), \textit{i.e.,} the interaction (\ref{hamiltonian}) is performed on atoms (6,7) and the entangled state (\ref{state1-4}) is obtained for atoms (5,6,7,8) and modes $(a_1,a_2;b_1,b_2)$. Finally, by applying the projection operators which are performed with the states $\ket{0;0} \otimes \ket{\widetilde{3};\widetilde{1}} $, $\ket{0;0} \otimes \ket{\widetilde{1};\widetilde{3}} $ and $\ket{1;1} \otimes \ket{\widetilde{3};\widetilde{3}} $ related to the state of field modes ($a_1;b_1$) and the state of atomic pair $(6;7)$ on entangled state (\ref{state1-4}) related to atoms (5,6,7,8), the following entangled states are respectively achieved for atoms (5,8)
\begin{eqnarray}\label{state158}
    \ket{\psi(t)}^1_{5,8}&=&\frac{1}{\sqrt{P^1_{1,4}(t)} }(A_2(t)\ket{\widetilde{1},\widetilde{3}}+A_{10}(t)\ket{\widetilde{3},\widetilde{1}}),
 \end{eqnarray}
 \begin{eqnarray}\label{state258}
     \ket{\psi(t)}^2_{5,8}&=&\frac{1}{\sqrt{P^1_{1,4}(t)} }(A_{10}(t)\ket{\widetilde{1},\widetilde{3}}+A_{2}(t)\ket{\widetilde{3},\widetilde{1}}),
  \end{eqnarray}
 and
 \begin{eqnarray}\label{state358}
     \ket{\psi(t)}^3_{5,8}&=&\frac{1}{\sqrt{2} }(\ket{\widetilde{1},\widetilde{3}}+\ket{\widetilde{3},\widetilde{1}}).
  \end{eqnarray}
At last, two three-level atoms (4,5) are interacted in an optical cavity based on the following Hamiltonian
\begin{eqnarray}\label{hamiltonian2}
 \hat{H}_{(4,5)}&=&\sum^3_{i=1}\widetilde{\omega}_i\hat{\sigma}_{ii}^4+\sum^3_{i=1}\widetilde{\omega}^{'}_i\hat{\sigma}_{ii}^5+\sum_{j=1,2}\Omega_j\hat{a}_j^{\dagger}\hat{a}_j\\\nonumber
   &+&\lambda_1(\hat{a}_1\hat{\sigma}_{13}^4+\hat{a}_1^{\dagger}\hat{\sigma}_{31}^4)+\lambda_2(\hat{a}_2\hat{\sigma}_{23}^4+\hat{a}_2^{\dagger}\hat{\sigma}_{32}^4)\\\nonumber
     &+&\lambda^{'}_1(\hat{a}_1\hat{\sigma}_{13}^5+\hat{a}_1^{\dagger}\hat{\sigma}_{31}^5)+\lambda^{'}_2(\hat{a}_2\hat{\sigma}_{23}^5+\hat{a}_2^{\dagger}\hat{\sigma}_{32}^5).
  \end{eqnarray}
  Then, via considering the vacuum states for optical modes the following effective Hamiltonian can be achieved \cite{James2007,Gamel2010}
 \begin{eqnarray}\label{effectivehamiltonian2}
 \hat{H}^{\mathrm{eff}}_{(4,5)}&=&\frac{\lambda^2_1}{\omega_M}\left[ \sum_{i=4,5}\hat{\sigma}_{11}^i+\left( \hat{\sigma}_{13}^4\hat{\sigma}_{31}^5+\hat{\sigma}_{13}^5\hat{\sigma}_{31}^4\right)\right] \\\nonumber
   &+&\frac{\lambda^2_2}{\omega_M}\left[ \sum_{i=4,5} \hat{\sigma}_{22}^i+\left( \hat{\sigma}_{23}^4\hat{\sigma}_{32}^5+\hat{\sigma}_{23}^5\hat{\sigma}_{32}^4\right)\right].
 \end{eqnarray}
Using the effective Hamiltonian (\ref{effectivehamiltonian2}), the time-dependent Schr$\mathrm{\ddot{o}}$dinger equation $i\frac{\partial}{\partial \tau}\ket{\psi(\tau)}^i_{(1,4,5,8)}=\hat{H}^{\mathrm{eff}}_{(4,5)}\ket{\psi(\tau)}^i_{(1,4,5,8)}$ where $i=1,2,3,4$ refer to the different initial states $\ket{\psi(t)}^{1}_{1,4}\otimes\ket{\psi(t)}^{2}_{5,8}$, $\ket{\psi(t)}^{2}_{1,4}\otimes\ket{\psi(t)}^{1}_{5,8}$, $\ket{\psi(t)}^{1}_{1,4}\otimes\ket{\psi(t)}^{1}_{5,8}$ and $\ket{\psi(t)}^{2}_{1,4}\otimes\ket{\psi(t)}^{2}_{5,8}$, (because of unsatisfactory values of success probability of states (\ref{state314}), (\ref{state358}), these states are set aside), the entangled state of atoms (1,4,5,8) is achieved as
 \begin{eqnarray}\label{state1-8}
  \ket{\psi(\tau)}^i_{(1,4,5,8)}&=&B^i_1(\tau)\ket{\widetilde{1},\widetilde{3};\widetilde{1},\widetilde{3}}+B^i_2(\tau)\ket{\widetilde{1},\widetilde{1};\widetilde{3},\widetilde{3}}\\\nonumber
    &+&B^i_3(\tau)\ket{\widetilde{1},\widetilde{3};\widetilde{3},\widetilde{1}}+B^i_4(\tau)\ket{\widetilde{3},\widetilde{1};\widetilde{1},\widetilde{3}}\\\nonumber
      &+&B^i_5(\tau)\ket{\widetilde{3},\widetilde{1};\widetilde{3},\widetilde{1}}+B^i_6(\tau)\ket{\widetilde{3},\widetilde{3};\widetilde{1},\widetilde{1}},
 \end{eqnarray}
where $\tau$ ($\tau>t$) is the interaction time between atoms (4,5) and for\\
 \begin{itemize}
  \item{\textit{i}=1, \textit{i.e.,} when the initial state is $\ket{\psi(t)}^{1}_{1,4}\otimes\ket{\psi(t)}^{2}_{5,8}$}
 \begin{eqnarray}\label{coef}
B^1_1(\tau)&=&B^1_5(\tau)=\frac{A_2(t)A_{10}(t)}{2P^1_{1,4}(t)}\left( 1+e^{-i\frac{2\lambda_1^2}{\omega_M}(\tau-t)}\right),\\\nonumber B^1_2(\tau)&=&B^1_6(\tau)=-\frac{A_2(t)A_{10}(t)}{2P^1_{1,4}(t)}\left( 1-e^{-i\frac{2\lambda_1^2}{\omega_M}(\tau-t)}\right) ,\\\nonumber
B^1_3(\tau)&=&\frac{A^2_2(t)}{P^1_{1,4}(t)},\qquad B^1_4(\tau)=\frac{A^2_{10}(t)}{P^1_{1,4}(t)}e^{-i\frac{2\lambda_1^2}{\omega_M}(\tau-t)},
 \end{eqnarray}
  \item{\textit{i}=2, \textit{i.e.,} when the initial state is $\ket{\psi(t)}^{2}_{1,4}\otimes\ket{\psi(t)}^{1}_{5,8}$}
  \begin{eqnarray}\label{coef}
 B^2_1(\tau)&=&B^2_5(\tau)=\frac{A_2(t)A_{10}(t)}{2P^1_{1,4}(t)}\left( 1+e^{-i\frac{2\lambda_1^2}{\omega_M}(\tau-t)}\right),\\\nonumber B^2_2(\tau)&=&B^2_6(\tau)=-\frac{A_2(t)A_{10}(t)}{2P^1_{1,4}(t)}\left( 1-e^{-i\frac{2\lambda_1^2}{\omega_M}(\tau-t)}\right) ,\\\nonumber
 B^2_3(\tau)&=&\frac{A^2_{10}(t)}{P^1_{1,4}(t)},\qquad B^2_4(\tau)=\frac{A^2_{2}(t)}{P^1_{1,4}(t)}e^{-i\frac{2\lambda_1^2}{\omega_M}(\tau-t)},
  \end{eqnarray}
  \item{\textit{i}=3, \textit{i.e.,} when the initial state is $\ket{\psi(t)}^{1}_{1,4}\otimes\ket{\psi(t)}^{1}_{5,8}$}
    \begin{eqnarray}\label{coef}
   B^3_1(\tau)&=&\frac{A^2_2(t)}{2P^1_{1,4}(t)}\left( 1+e^{-i\frac{2\lambda_1^2}{\omega_M}(\tau-t)}\right),\\\nonumber B^3_2(\tau)&=&-\frac{A^2_2(t)}{2P^1_{1,4}(t)}\left( 1-e^{-i\frac{2\lambda_1^2}{\omega_M}(\tau-t)}\right) ,\\\nonumber
   B^3_3(\tau)&=&\frac{A_{2}(t)A_{10}(t)}{P^1_{1,4}(t)},\qquad B^3_4(\tau)=\frac{A_{2}(t)A_{10}(t)}{P^1_{1,4}(t)}e^{-i\frac{2\lambda_1^2}{\omega_M}(\tau-t)},\\\nonumber
   B^3_5(\tau)&=&\frac{A^2_{10}(t)}{2P^1_{1,4}(t)}\left( 1+e^{-i\frac{2\lambda_1^2}{\omega_M}(\tau-t)}\right),\\\nonumber
   B^3_6(\tau)&=&-\frac{A^2_{10}(t)}{2P^1_{1,4}(t)}\left( 1-e^{-i\frac{2\lambda_1^2}{\omega_M}(\tau-t)}\right),
    \end{eqnarray}
  \item{\textit{i}=4, \textit{i.e.,} when the initial state is $\ket{\psi(t)}^{2}_{1,4}\otimes\ket{\psi(t)}^{2}_{5,8}$}
        \begin{eqnarray}\label{coef}
       B^4_1(\tau)&=&\frac{A^2_{10}(t)}{2P^1_{1,4}(t)}\left( 1+e^{-i\frac{2\lambda_1^2}{\omega_M}(\tau-t)}\right),\\\nonumber B^4_2(\tau)&=&-\frac{A^2_{10}(t)}{2P^1_{1,4}(t)}\left( 1-e^{-i\frac{2\lambda_1^2}{\omega_M}(\tau-t)}\right) ,\\\nonumber
       B^4_3(\tau)&=&\frac{A_{2}(t)A_{10}(t)}{P^1_{1,4}(t)},\qquad B^4_4(\tau)=\frac{A_{2}(t)A_{10}(t)}{P^1_{1,4}(t)}e^{-i\frac{2\lambda_1^2}{\omega_M}(\tau-t)},\\\nonumber
       B^4_5(\tau)&=&\frac{A^2_{2}(t)}{2P^1_{1,4}(t)}\left( 1+e^{-i\frac{2\lambda_1^2}{\omega_M}(\tau-t)}\right),\\\nonumber
       B^4_6(\tau)&=&-\frac{A^2_{2}(t)}{2P^1_{1,4}(t)}\left( 1-e^{-i\frac{2\lambda_1^2}{\omega_M}(\tau-t)}\right).
        \end{eqnarray}
        \end{itemize}
 Now, measuring the state $\ket{\widetilde{1};\widetilde{3}}$ and $\ket{\widetilde{3};\widetilde{1}}$ related to atoms (4,5) on state (\ref{state1-8}), the states of target atoms (1,8) are respectively converted to entangled states
\begin{eqnarray}\label{state18}
    \ket{\psi(\tau)}^i_{(1,8)}&=&\frac{1}{\sqrt{P^i_{1,8}(\tau)} }(B^i_2(\tau)\ket{\widetilde{1},\widetilde{3}}+B^i_5(\tau)\ket{\widetilde{3},\widetilde{1}}),
 \end{eqnarray}
  with the entropy and success probability as
    \begin{eqnarray}\label{ent18}
      E^{i}_{1,8}(\tau)&=&1-\frac{\left| B^i_2(\tau)\right| ^4+\left| B^i_5(\tau)\right|^4}{(P^i_{1,8}(\tau))^2},
        \end{eqnarray}
  \begin{eqnarray}\label{suc18}
 P^{i}_{1,8}(\tau)&=&\left| B^i_2(\tau)\right| ^2+\left| B^i_5(\tau)\right|^2 ,
   \end{eqnarray}
     and
\begin{eqnarray}\label{state218}
\ket{\psi^{'}(\tau)}^i_{(1,8)}&=&\frac{1}{\sqrt{P^{'i}_{1,8}(\tau)} }(B^i_1(\tau)\ket{\widetilde{1},\widetilde{3}}+B^i_6(\tau)\ket{\widetilde{3},\widetilde{1}}),
 \end{eqnarray}
 with the entropy and success probability as
   \begin{eqnarray}\label{ent218}
     E^{'i}_{1,8}(\tau)&=&1-\frac{\left| B^i_1(\tau)\right| ^4+\left| B^i_6(\tau)\right|^4}{(P^{'i}_{1,8}(\tau))^2},
       \end{eqnarray}
 \begin{eqnarray}\label{suc218}
 P^{'i}_{1,8}(\tau)&=&\left| B^i_1(\tau)\right| ^2+\left| B^i_6(\tau)\right|^2.
   \end{eqnarray}
  Notice that $E^{j}_{1,8}(\tau)=E^{'j}_{1,8}(\tau)$ and $ P^{j}_{1,8}(\tau) =P^{'j}_{1,8}(\tau)$, where $j=1,2$ and $E^{1}_{1,8}(\tau)=E^{2}_{1,8}(\tau)$, $ P^{1}_{1,8}(\tau) =P^{2}_{1,8}(\tau)$, also, $E^{3}_{1,8}(\tau)=E^{'4}_{1,8}(\tau)$, $ P^{3}_{1,8}(\tau) =P^{'4}_{1,8}(\tau)$, $E^{4}_{1,8}(\tau)=E^{'3}_{1,8}(\tau)$ and $ P^{4}_{1,8}(\tau) =P^{'3}_{1,8}(\tau)$. In the next section we analyze the entanglement and success probability of entangled states of atomic pairs (1,4) (or (5,8)) and (1,8).
 
 \section{Results and discussion} \label{sec.results}

    \begin{figure}[H]
         \centering
         \subfigure[\label{fig.Fig2a} \ $E^1_{1,4}(t)$]{\includegraphics[width=0.33\textwidth]{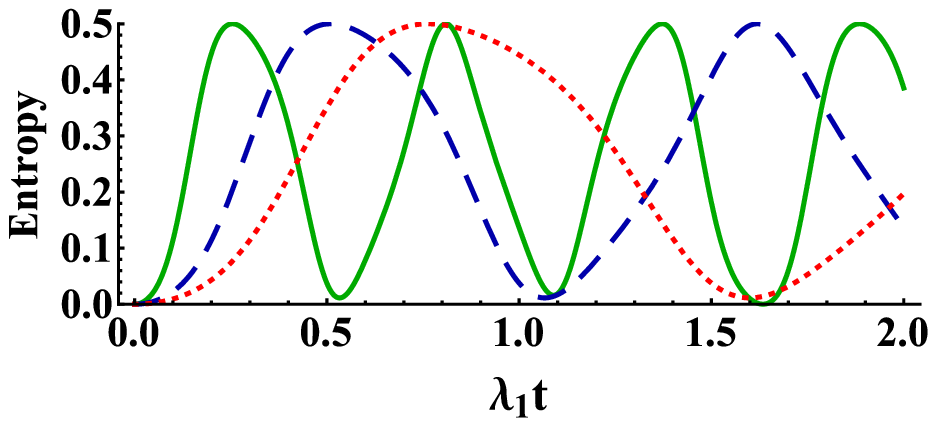}}
         \hspace{0.05\textwidth}
         \subfigure[\label{fig.Fig2b} \ $P^1_{1,4}(t)$]{\includegraphics[width=0.33\textwidth]{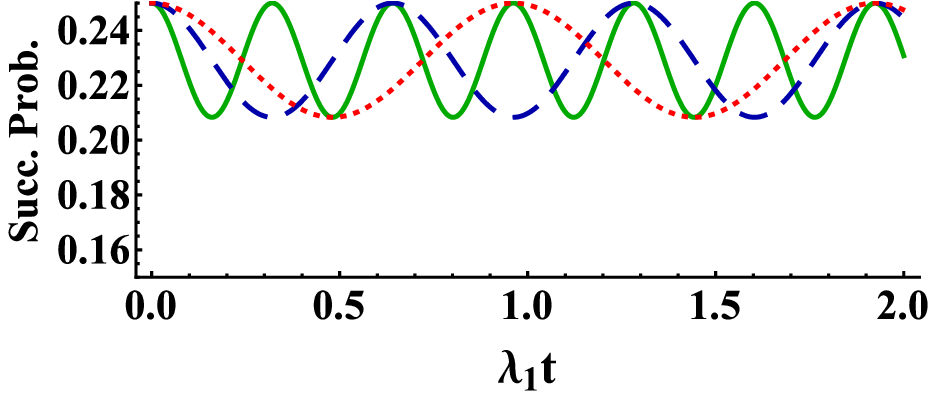}}
         \caption{\label{fig.fig2} {\it The effect of mechanical frequency, $\omega_M$, on the evolution of}: (a) entropy (Eq. (\ref{ent14})) and (b) success probability (Eq. (\ref{suc14})) for $\omega_M/\lambda_1=0.5$ (solid green line), $\omega_M/\lambda_1=1$ (dashed blue line), $\omega_M/\lambda_1=1.5$ (dotted red line) with $G/\lambda_1=2$.}
         \end{figure}
   \begin{figure}[H]
    \centering
    \subfigure[\label{fig.Fig3a} \ $E^1_{1,4}(t)$]{\includegraphics[width=0.33\textwidth]{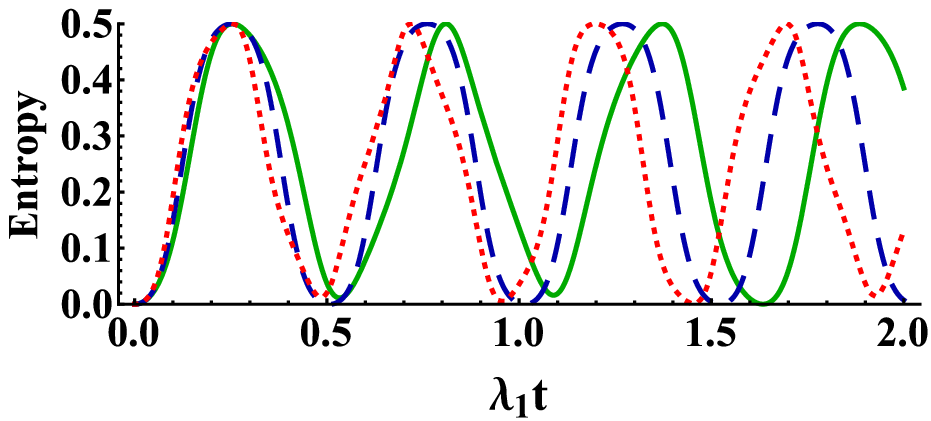}}
    \hspace{0.05\textwidth}
    \subfigure[\label{fig.Fig3b} \  $P^1_{1,4}(t)$]{\includegraphics[width=0.33\textwidth]{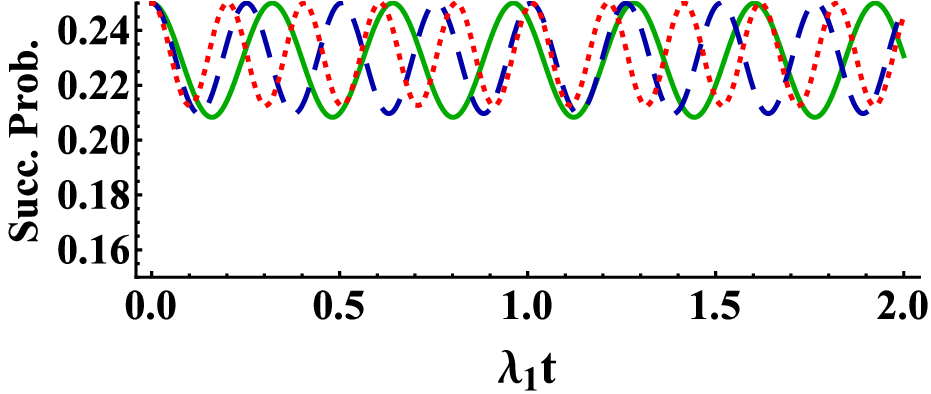}}
    \caption{\label{fig.fig3} {\it The effect of optomechanical coupling strength to the field modes, $G$, on the evolution of}: (a) entropy (Eq. (\ref{ent14})) and (b) success probability (Eq. (\ref{suc14})) for $G/\lambda_1=2$ (solid green line), $G/\lambda_1=2.5$ (dashed blue line), $G/\lambda_1=3$ (dotted red line) with $\omega_M/\lambda_1=0.5$.}
    \end{figure}
    
       \begin{figure}[H]
       \centering
       \subfigure[\label{fig.Fig4a} \ $P^{2}_{1,4}(t)$]{\includegraphics[width=0.33\textwidth]{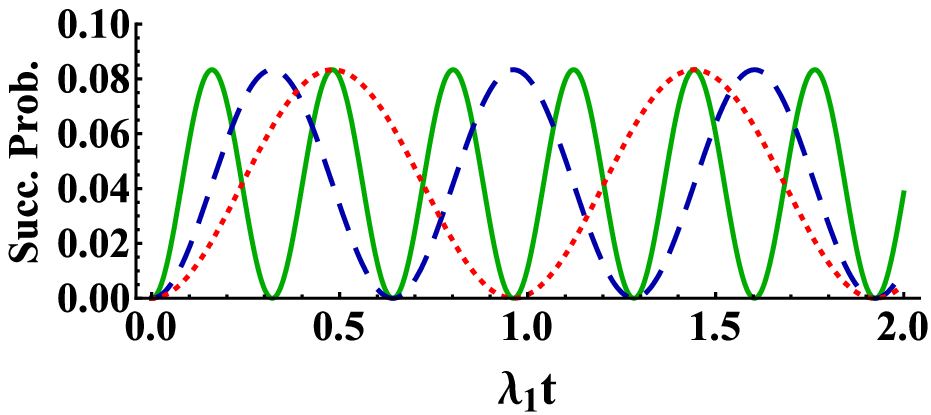}}
       \hspace{0.05\textwidth}
       \subfigure[\label{fig.Fig4b} \ $P^{2}_{1,4}(t)$]{\includegraphics[width=0.33\textwidth]{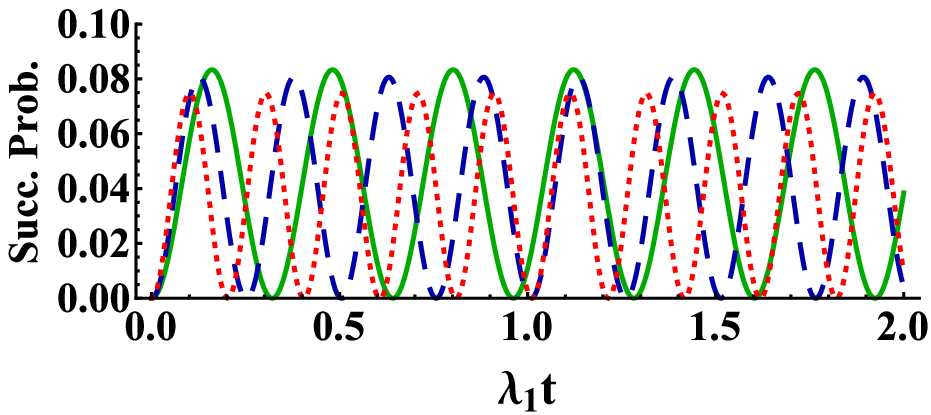}}
       \caption{\label{fig.fig4} {\it The effects of mechanical frequency, $\omega_M$, and optomechanical coupling strength to the field modes, $G$, on the evolution of}: success probability (Eq. (\ref{suc214})) for (a) $\omega_M/\lambda_1=0.5$ (solid green line), $\omega_M/\lambda_1=1$ (dashed blue line), $\omega_M/\lambda_1=1.5$ (dotted red line) with $G/\lambda_1=2$ and (b) $G/\lambda_1=2$ (solid green line), $G/\lambda_1=2.5$ (dashed blue line), $G/\lambda_1=3$ (dotted red line) with $\omega_M/\lambda_1=0.5$.}
       \end{figure}
     \begin{figure}[H]
       \centering
       \subfigure[\label{fig.Fig5a} \  $E^{1}_{1,8}(\tau)=E^{2}_{1,8}(\tau)=E^{'1}_{1,8}(\tau)=E^{'2}_{1,8}(\tau)$]{\includegraphics[width=0.33\textwidth]{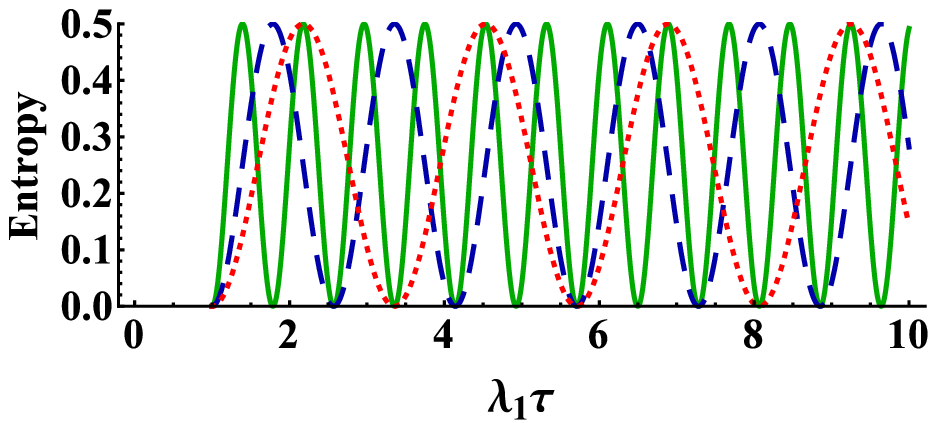}}
       \hspace{0.05\textwidth}
       \subfigure[\label{fig.Fig5b} \  $P^{1}_{1,8}(\tau)=P^{2}_{1,8}(\tau)=P^{'1}_{1,8}(\tau)=P^{'2}_{1,8}(\tau)$]{\includegraphics[width=0.33\textwidth]{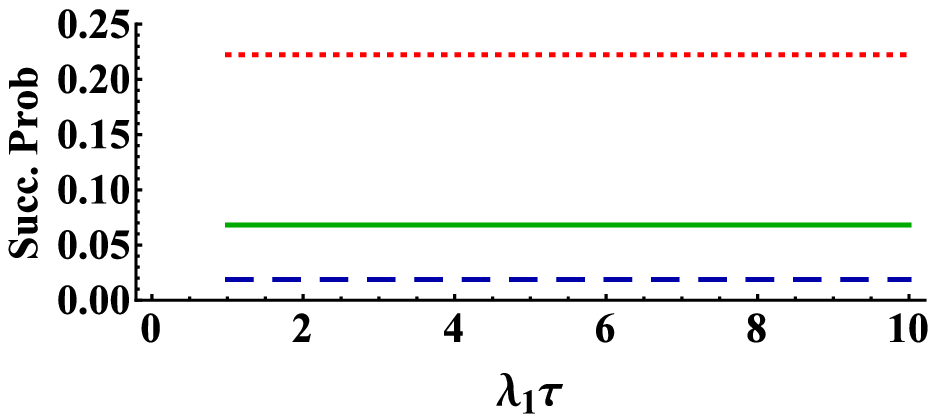}}
        \hspace{0.05\textwidth}
             \subfigure[\label{fig.Fig5c} \  $E^{3}_{1,8}(\tau)=E^{'4}_{1,8}(\tau)$]{\includegraphics[width=0.33\textwidth]{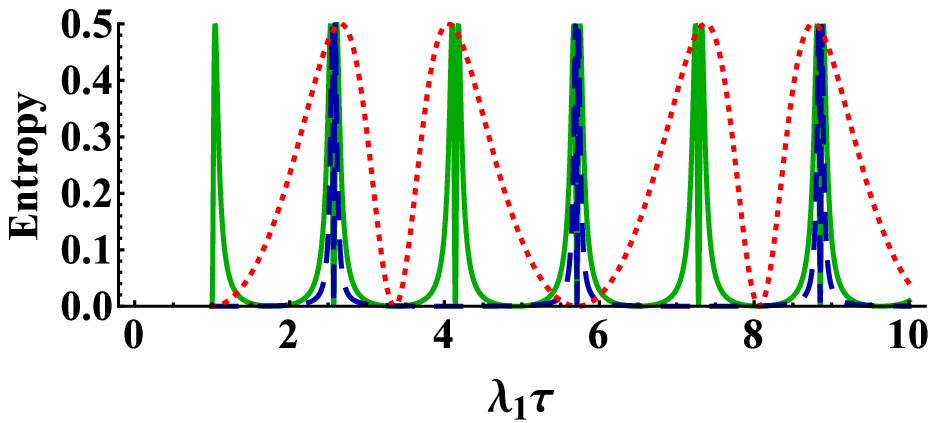}}
              \hspace{0.05\textwidth}
                   \subfigure[\label{fig.Fig5d} \ $P^{3}_{1,8}(\tau)=P^{'4}_{1,8}(\tau)$]{\includegraphics[width=0.33\textwidth]{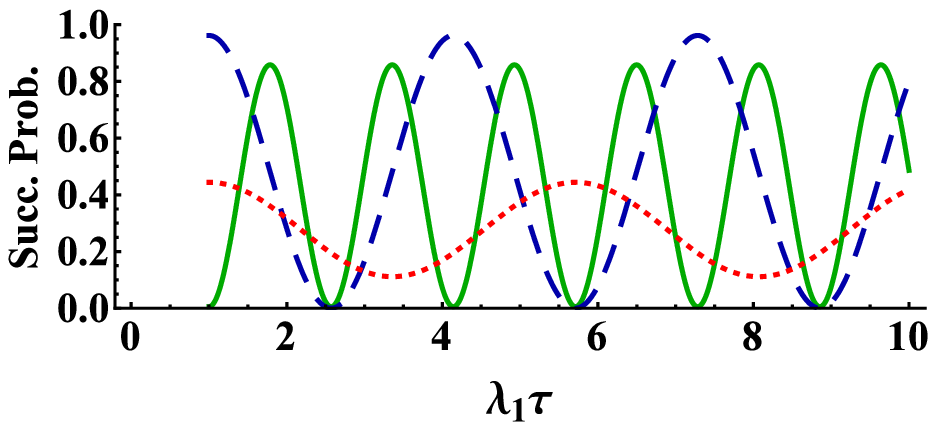}}
                    \hspace{0.05\textwidth}
                         \subfigure[\label{fig.Fig5e} \ $E^{4}_{1,8}(\tau)=E^{'3}_{1,8}(\tau)$]{\includegraphics[width=0.33\textwidth]{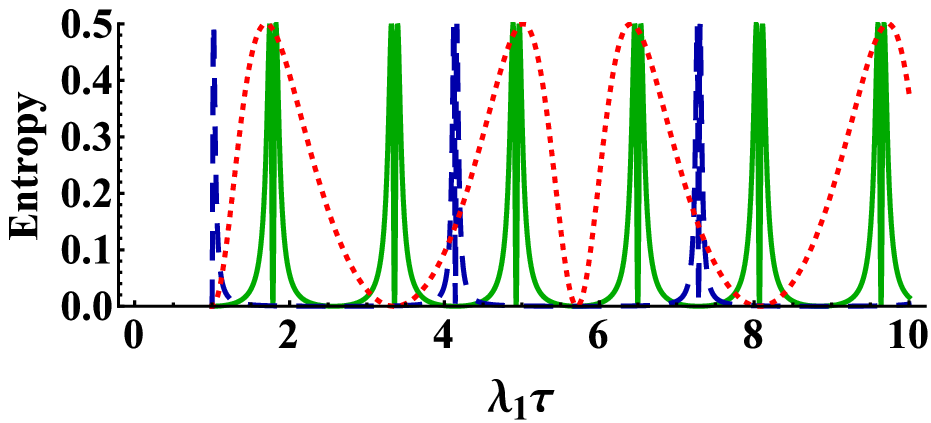}}
                          \hspace{0.05\textwidth}
                               \subfigure[\label{fig.Fig5f} \ $P^{4}_{1,8}(\tau)=P^{'3}_{1,8}(\tau)$]{\includegraphics[width=0.33\textwidth]{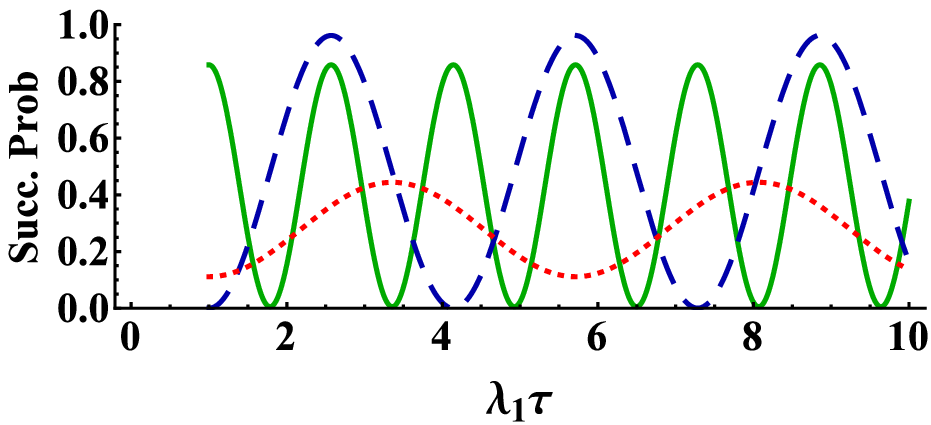}}
                                     \caption{\label{fig.fig5} {\it The effect of mechanical frequency, $\omega_M$, on the evolution of}: (a) entropy (Eqs. (\ref{ent18}), (\ref{ent218})) and (b) success probability (Eqs. (\ref{suc18}), (\ref{suc218})) for $\omega_M/\lambda_1=0.5$ (solid green line), $\omega_M/\lambda_1=1$ (dashed blue line), $\omega_M/\lambda_1=1.5$ (dotted red line) with $G/\lambda_1=2$ and $\lambda_1 t=1$.}
       \end{figure}
        \begin{figure}[H]
             \centering
             \subfigure[\label{fig.Fig6a} \ $E^{1}_{1,8}(\tau)=E^{2}_{1,8}(\tau)=E^{'1}_{1,8}(\tau)=E^{'2}_{1,8}(\tau)$]{\includegraphics[width=0.33\textwidth]{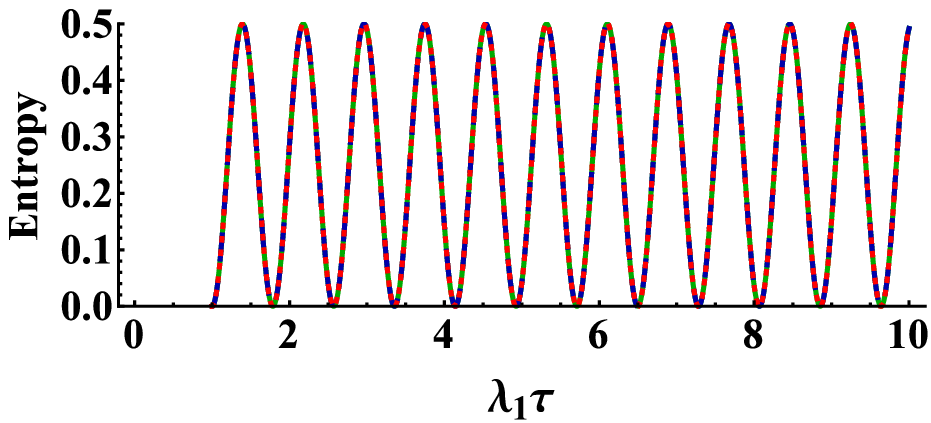}}
             \hspace{0.05\textwidth}
             \subfigure[\label{fig.Fig6b} \ $P^{1}_{1,8}(\tau)=P^{2}_{1,8}(\tau)=P^{'1}_{1,8}(\tau)=P^{'2}_{1,8}(\tau)$]{\includegraphics[width=0.33\textwidth]{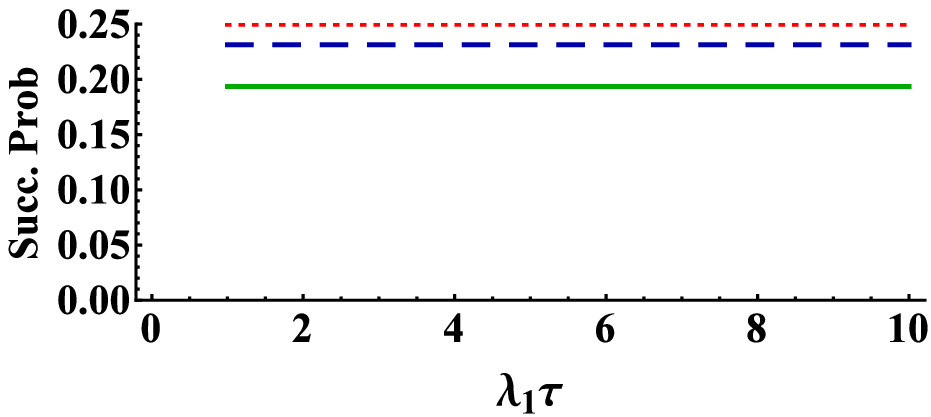}}
              \hspace{0.05\textwidth}
                   \subfigure[\label{fig.Fig6c} \ $E^{3}_{1,8}(\tau)=E^{'4}_{1,8}(\tau)$]{\includegraphics[width=0.33\textwidth]{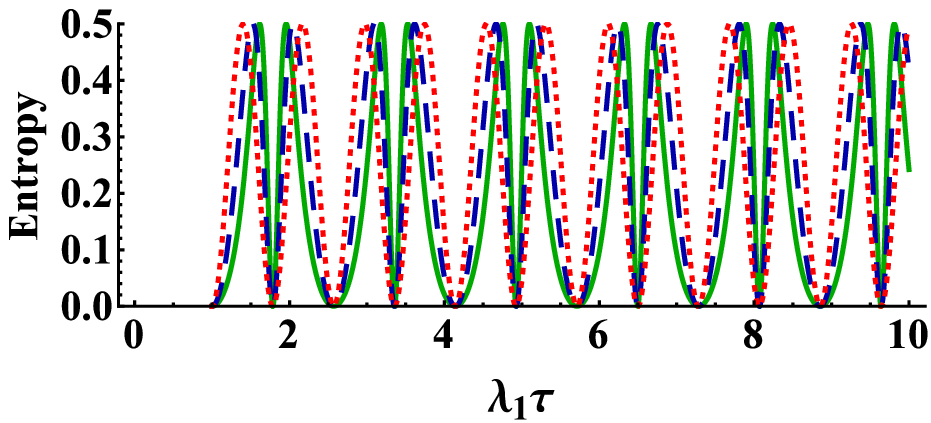}}
                    \hspace{0.05\textwidth}
                         \subfigure[\label{fig.Fig6d} \ $P^{3}_{1,8}(\tau)=P^{'4}_{1,8}(\tau)$]{\includegraphics[width=0.33\textwidth]{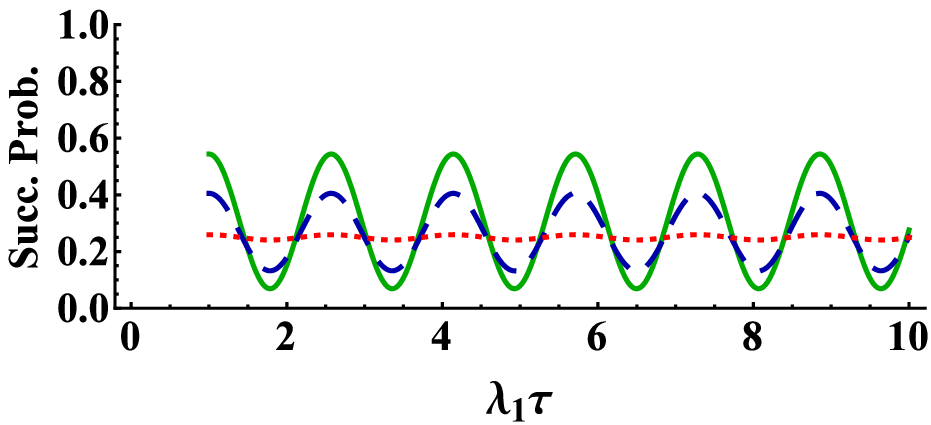}}
                          \hspace{0.05\textwidth}
                               \subfigure[\label{fig.Fig6e} \ $E^{4}_{1,8}(\tau)=E^{'3}_{1,8}(\tau)$]{\includegraphics[width=0.33\textwidth]{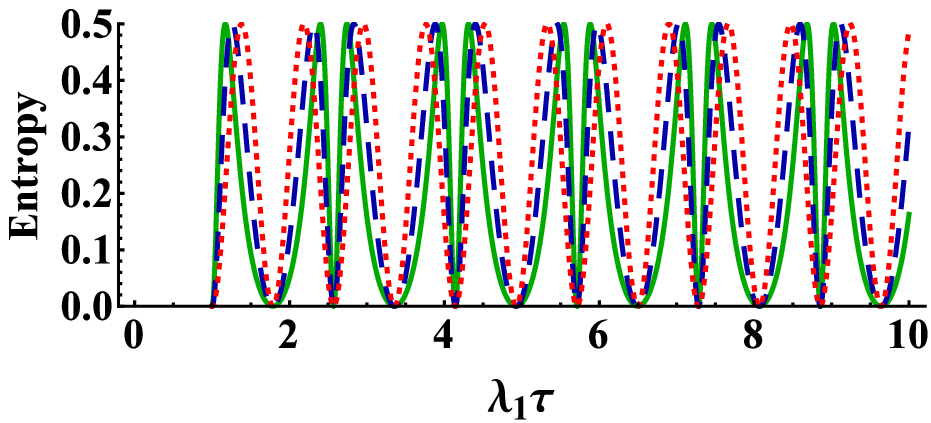}}
                                \hspace{0.05\textwidth}
                                     \subfigure[\label{fig.Fig6f} \ $P^{4}_{1,8}(\tau)=P^{'3}_{1,8}(\tau)$]{\includegraphics[width=0.33\textwidth]{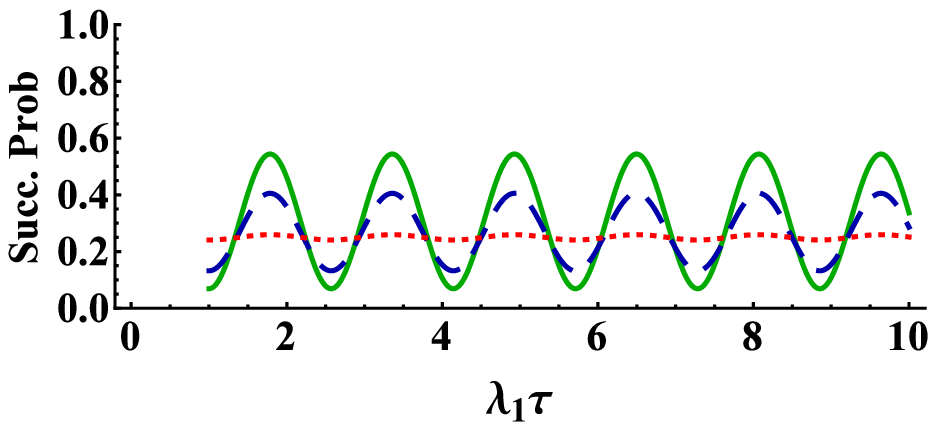}}
                                                 \caption{\label{fig.fig6} {\it The effect of optomechanical coupling strength to the field modes, $G$, on the evolution of}: (a) entropy (Eqs. (\ref{ent18}), (\ref{ent218})) and (b) success probability (Eqs. (\ref{suc18}), (\ref{suc218})) for $G/\lambda_1=0.5$ (solid green line), $G/\lambda_1=0.7$ (dashed blue line), $G/\lambda_1=0.9$ (dotted red line) with $\omega_M/\lambda_1=0.5$ and $\lambda_1 t=1$.}
             \end{figure}

 The aim of this section is to analyze  the specific features of our introduced quantum repeater.  
 As is usual, in such protocols the success probability and the amount of entanglement are the key points which should be evaluated.    
 To achieve the purpose, we have plotted the figures of entropy and success probability of pairs (1,4) (or (5,8)) and (1,8) and the effects of different parameters are considered on these quantities.
 In figures \ref{fig.fig2} and \ref{fig.fig3} the effects of mechanical frequency and optomechanical coupling strength to the field modes on the entropy and success probability of pair (1,4) (or (5,8)) related to states (\ref{state114}) and (\ref{state214}) are respectively considered. From figures \ref{fig.Fig2a} and \ref{fig.Fig2b} (\ref{fig.Fig3a} and \ref{fig.Fig3b}) it can be seen that, the time periods of entropy and success probability are increased (decreased) by increasing $\omega_M$ ($G$).
 In figures \ref{fig.Fig4a} and \ref{fig.Fig4b} the effects of mechanical frequency and optomechanical coupling strength on the success probability of entangled state (\ref{state314}) are considered. From these figures we can see that the time period of success probability is increased (decreased) by increasing $\omega_M$ ($G$).
 In figures \ref{fig.fig5} and \ref{fig.fig6} the effects of mechanical frequency and optomechanical coupling strength to the field modes on the entropy and success probability of target atoms (1,8) are respectively considered. In figures \ref{fig.Fig5a}, \ref{fig.Fig5c} and \ref{fig.Fig5e} by increasing the mechanical frequency, the time period of entropy is increased. We can see from figure \ref{fig.Fig5b} that the success probability independent of $\tau$ is increased by increasing $\omega_M$. Also, in figures \ref{fig.Fig5d} and \ref{fig.Fig5f} the time periods of success probability are increased by increasing $\omega_M$ and the maxima of success probability in these figures reach to $1$ for $\omega_M=\lambda_1$.\\
  In figure \ref{fig.Fig6a} (\ref{fig.Fig6b}) we find that the entropy (success probability) is independent of $G$ ($\tau$). In figures \ref{fig.Fig6c} and \ref{fig.Fig6e}, the entropy is approximately unchanged by increasing $G$. Also, in figures \ref{fig.Fig6d} and \ref{fig.Fig6f}, the maxima of success probability are decreased by increasing $G$.

\section{Summary and conclusions} \label{sec.Conclusion}
In this paper we considered entanglement distribution to two three-level distant atoms based on quantum repeater protocol 
using the optical cavity QED method as well as optomechanical systems. 
In our proposal six atoms are inserted between the two target atoms. 
Preparing the atomic pairs $(i,i+1)$ where $i=1,3,5,7$ in some entangled states, 
the separable atoms (1,4) and (5,8) were converted to entangled states by performing interaction between atoms 
(2,3) as well as (6,7) in two separate optomechanical cavities. 
At last, by performing the interaction between two atoms (4,5) in an optical cavity, the target atoms (1,8) become entangled. Our
numerical results show that, the time periods of entropy and success probability of pairs (1,4) and (5,8) are increased (decreased) by increasing $\omega_M$ ($G$). Also, the time periods of entropy and success probability of target atoms (1,8) are increased by increasing $\omega_M$. But, the time period of entanglement of atoms (1,8) is unchanged by increasing $G$, also, the maxima of success probability are decreased by increasing $G$. In our consideration, using QED-optomechanical hybrid systems, the acceptable success probability related to target distributed entangled state for some cases has been obtained independent of interaction time. It is noticeable that satisfactorily degree of entanglement $(0.5)$ and acceptable value of success probability are achievable by justifying the parameters of optomechanics and cavity QED used methods (see Refs. \cite{Ralph2001,Knill2001,Bergou2005,Scheel2006}).

 

\begin{thebibliography}{999}

\bibitem{Briegel1998} H.-J. Briegel, W. Dür, J. I. Cirac, and P. Zoller, “Quantum repeaters: the role of imperfect local operations in quantum communication,” \textit{Phys. Rev. Lett.} \textbf{81}, 5932 (1998).

\bibitem{Yi2019} X.-F. Yi, P. Xu, Q. Yao, and X. Quan, “Quantum repeater without bell
measurements in double-quantum-dot systems,” \textit{Quantum Inf. Process.}
\textbf{18}, 82 (2019).

\bibitem{Uphoff2016} M. Uphoff, M. Brekenfeld, G. Rempe, and S. Ritter, “An integrated
quantum repeater at telecom wavelength with single atoms in optical
fiber cavities,” \textit{Appl. Phys. B} \textbf{122}, 46 (2016).

\bibitem{Bernad2013} J. Bernad, H. Frydrych, and G. Alber, “Centre-of-mass motion-induced
decoherence and entanglement generation in a hybrid quantum repeater,”
\textit{J. Phys. B: At. Mol. Opt. Phys.} \textbf{46}, 235501 (2013).

\bibitem{Ghasemi20193} M. Ghasemi and M. K. Tavassoly, “Toward a quantum repeater protocol
based on the coherent state approach,” \textit{Laser Phys.} \textbf{29}, 085202 (2019).

\bibitem{Furrer2018} F. Furrer and W. J. Munro, “Repeaters for continuous-variable quantum
communication,” \textit{Phys. Rev. A} \textbf{98}, 032335 (2018).

\bibitem{Li2016} T. Li, G.-J. Yang, and F.-G. Deng, “Heralded quantum repeater for a
quantum communication network based on quantum dots embedded in
optical microcavities,” \textit{Phys. Rev. A} \textbf{93}, 012302 (2016).

\bibitem{Ghasemi2018} M. Ghasemi and M. K. Tavassoly, “Quantum repeater protocol in mixed
single-and two-mode tavis-cummings models,” \textit{EPL (Europhysics Letters)}
\textbf{123}, 24002 (2018).

\bibitem{Ghasemi2019} M. Ghasemi and M. K. Tavassoly, “Dissipative quantum repeater,” \textit{Quantum
Inf. Process.} \textbf{18}, 113 (2019).

\bibitem{Ghasemi20192} M. Ghasemi and M. K. Tavassoly, “Quantum repeater using three-level
atomic states in the presence of dissipation: stability of entanglement,” \textit{J.
Phys. B: At. Mol. Opt. Phys.} \textbf{52}, 085502 (2019).

\bibitem{Azuma2017} K. Azuma and G. Kato, “Aggregating quantum repeaters for the quantum
internet,” \textit{Phys. Rev. A} \textbf{96}, 032332 (2017).

\bibitem{Behera2019} B. K. Behera, S. Seth, A. Das, and P. K. Panigrahi, “Demonstration
of entanglement purification and swapping protocol to design quantum
repeater in ibm quantum computer,” \textit{Quantum Inf. Process.} \textbf{18}, 108
(2019).

\bibitem{Ghasemi2016} M. Ghasemi and M. K. Tavassoly, “Entanglement swapping to a qutritqutrit
atomic system in the presence of kerr medium and detuning
parameter,” \textit{Eur. Phys. J. Plus} \textbf{131}, 297 (2016).

\bibitem{Ghasemi2017} M. Ghasemi, M. K. Tavassoly, and A. Nourmandipour, “Dissipative
entanglement swapping in the presence of detuning and kerr medium:
Bell state measurement method,” \textit{Eur. Phys. J. Plus} \textbf{132}, 531 (2017).

\bibitem{Nourmandipour2016} A. Nourmandipour and M. K. Tavassoly, “Entanglement swapping between
dissipative systems,” \textit{Phys. Rev. A} \textbf{94}, 022339 (2016).

\bibitem{Liao2011} Q. Liao, G. Fang, Y. Wang, M. Ahmad, and S. Liu, “Entanglement
swapping in two independent jaynes-cummings models,” \textit{Eur. Phys. J. D}
\textbf{61}, 475–479 (2011).


\bibitem{Pakniat2017} R. Pakniat, M. K. Tavassoly, and M. H. Zandi, “Entanglement swapping
and teleportation based on cavity qed method using the nonlinear atom–
field interaction: Cavities with a hybrid of coherent and number states,”
\textit{Opt. Commun.} \textbf{382}, 381–385 (2017).

\bibitem{Soltani2017} M. Soltani, M. K. Tavassoly, and R. Pakniat, “The influence of excitation
number of photon-added coherent state field on the entanglement
swapping process,” \textit{Int. J. Mod. Phys. B} \textbf{31}, 1750198 (2017).

\bibitem{Xue2006} Z.-Y. Xue, M. Yang, Y.-M. Yi, and Z.-L. Cao, “Teleportation for atomic
entangled state by entanglement swapping with separate measurements
in cavity qed,” \textit{Opt. Commun.} \textbf{258}, 315–320 (2006).

\bibitem{Yang2005} M. Yang, W. Song, and Z.-L. Cao, “Entanglement swapping without
joint measurement,” \textit{Phys. Rev. A} \textbf{71}, 034312 (2005).

\bibitem{Yang2006} M. Yang and Z.-L. Cao, “Scheme for bell-state-measurement-free quantum
teleportation,” \textit{Int. J. Quantum Inf.} \textbf{4}, 341–346 (2006).

\bibitem{Osnaghi2001} S. Osnaghi, P. Bertet, A. Auffeves, P. Maioli, M. Brune, J.-M. Raimond,
and S. Haroche, “Coherent control of an atomic collision in a cavity,”
\textit{Phys. Rev. Lett.} \textbf{87}, 037902 (2001).

\bibitem{Cardoso2009} A. D. dSouza, W. B. Cardoso, A. T. Avelar, and B. Baseia, “Entanglement
swapping in the two-photon Jaynes–Cummings model,” \textit{Phys. Scr.}
\textbf{80}, 065009 (2009).

\bibitem{Giang2010} W.-C. Qiang, W. Cardoso, and X.-H. Zhang, “The entropy of entangled
three-level atoms interacting with entangled cavity fields: Entanglement
swapping,” \textit{Physica A: Stat. Mech. Appl.} \textbf{389}, 5109–5115 (2010).

\bibitem{Xiu2007} L. Xiu, L. Hong-Cai, Y. Rong-Can, and H. Zhi-Ping, “Entanglement
swapping without joint measurement via a l-type atom interacting with
bimodal cavity field,” \textit{Chin. Phys.} \textbf{16}, 919 (2007).

\bibitem{Jaynes1963} E. T. Jaynes and F. W. Cummings, “Comparison of quantum and semiclassical
radiation theories with application to the beam maser,” \textit{Proc.
IEEE} \textbf{51}, 89–109 (1963).

\bibitem{Bashkirov2006} E. K. Bashkirov, “Dynamics of the two-atom jaynes-cummings model
with nondegenerate two-photon transitions,” \textit{Laser Phys.} \textbf{16}, 1218–1226
(2006).

\bibitem{Gerry1992} C. C. Gerry and R. F. Welch, “Dynamics of a two-mode two-photon
jaynes-cummings model interacting with correlated su (1, 1) coherent
states,” \textit{J. Opt. Soc. Am. B} \textbf{9}, 290–297 (1992).

\bibitem{Bartzis1991} V. Bartzis and N. Nayak, “Two-photon jaynes-cummings model,” \textit{J. Opt.
Soc. Am. B} \textbf{8}, 1779–1786 (1991).

\bibitem{Obada2004} A.-S. Obada and H. A. Hessian, “Entanglement generation and entropy
growth due to intrinsic decoherence in the jaynes-cummings model,” \textit{J.
Opt. Soc. Am. B} \textbf{21}, 1535–1542 (2004).

\bibitem{Hu1989} G. Hu and P. K. Aravind, “Dynamical symmetries of the vacuum field
jaynes-cummings model,” \textit{J. Opt. Soc. Am. B} \textbf{6}, 1757–1763 (1989).

\bibitem{Tavis1968} M. Tavis and F. W. Cummings, “Exact solution for an nmolecule—
radiation-field hamiltonian,” \textit{Phys. Rev.} \textbf{170}, 379 (1968).

\bibitem{Bashkirov2012} E. K. Bashkirov and M. S. Rusakova, “Entanglement for two-atom
tavis–cummings model with degenerate two-photon transitions in the
presence of the stark shift,” \textit{Optik-Int. J. Light Electron Opt.} \textbf{123}, 1694–
1699 (2012).

\bibitem{Nadiki2016} M. H. Nadiki and M. K. Tavassoly, “Collapse-revival in entanglement
and photon statistics: the interaction of a three-level atom with a twomode
quantized field in cavity optomechanics,” \textit{Laser Phys.} \textbf{26}, 125204
(2016).

\bibitem{Nadiki2018} M. H. Nadiki, M. K. Tavassoly, and N. Yazdanpanah, “A trapped ion in
an optomechanical system: entanglement dynamics,” \textit{Eur. Phys. J. D} \textbf{72},
110 (2018).

\bibitem{Caves1980} C. M. Caves, “Quantum-mechanical radiation-pressure fluctuations in
an interferometer,” \textit{Phys. Rev. Lett.} \textbf{45}, 75 (1980).

\bibitem{Corbitt2006} T. Corbitt, D. Ottaway, E. Innerhofer, J. Pelc, and N. Mavalvala, “Measurement
of radiation-pressure-induced optomechanical dynamics in a
suspended Fabry-Perot cavity,” \textit{Phys. Rev. A} \textbf{74}, 021802 (2006).

\bibitem{Huang2010} S. Huang and G. Agarwal, “Normal-mode splitting and antibunching in
Stokes and anti-Stokes processes in cavity optomechanics: radiationpressure-
induced four-wave-mixing cavity optomechanics,” \textit{Phys. Rev. A}
\textbf{81}, 033830 (2010).

\bibitem{Joshi2012} C. Joshi, J. Larson, M. Jonson, E. Andersson, and P. Öhberg, “Entanglement
of distant optomechanical systems,” \textit{Phys. Rev. A} \textbf{85}, 033805
(2012).

\bibitem{Barzanjeh2011} S. Barzanjeh, M. Naderi, and M. Soltanolkotabi, “Steady-state entanglement
and normal-mode splitting in an atom-assisted optomechanical
system with intensity-dependent coupling,” \textit{Phys. Rev. A} \textbf{84}, 063850
(2011).

\bibitem{Safavi2011} A. H. Safavi-Naeini, T. M. Alegre, J. Chan, M. Eichenfield, M. Winger,
Q. Lin, J. T. Hill, D. E. Chang, and O. Painter, “Electromagnetically
induced transparency and slow light with optomechanics,” \textit{Nature} \textbf{472},
69 (2011).

\bibitem{Verhagen2012} E. Verhagen, S. Deléglise, S. Weis, A. Schliesser, and T. J. Kippenberg,
“Quantum-coherent coupling of a mechanical oscillator to an optical
cavity mode,” \textit{Nature} \textbf{482}, 63 (2012).

\bibitem{Anetsberger2009} G. Anetsberger, O. Arcizet, Q. P. Unterreithmeier, R. Rivière,
A. Schliesser, E. M. Weig, J. P. Kotthaus, and T. J. Kippenberg, “Nearfield
cavity optomechanics with nanomechanical oscillators,” \textit{Nat. Phys.}
\textbf{5}, 909 (2009).

\bibitem{Zhang2017} J.-S. Zhang, W. Zeng, and A.-X. Chen, “Effects of cross-Kerr coupling
and parametric nonlinearity on normal mode splitting, cooling, and entanglement
in optomechanical systems,” \textit{Quantum Inf. Process.} \textbf{16}, 163
(2017).

\bibitem{Eghbali2017} M. Eghbali-Arani and V. Ameri, “Entanglement of two hybrid optomechanical
cavities composed of BEC atoms under Bell detection,” \textit{Quantum
Inf. Process.} \textbf{16}, 47 (2017).

\bibitem{Feng2016} Z.-B. Feng, H.-L. Wang, and R.-Y. Yan, “Quantum state transfer between
an optomechanical cavity and a diamond nuclear spin ensemble,”
\textit{Quantum Inf. Process.} \textbf{15}, 3151–3167 (2016).

\bibitem{Galland2014} C. Galland, N. Sangouard, N. Piro, N. Gisin, and T. J. Kippenberg,
“Heralded single-phonon preparation, storage, and readout in cavity optomechanics,” \textit{Phys. Rev. Lett.} \textbf{112}, 143602 (2014).

\bibitem{Vitali2007} D. Vitali, S. Gigan, A. Ferreira, H. Böhm, P. Tombesi, A. Guerreiro,
V. Vedral, A. Zeilinger, and M. Aspelmeyer, “Optomechanical entanglement
between a movable mirror and a cavity field,” \textit{Phys. Rev. Lett.} \textbf{98},
030405 (2007).

\bibitem{Aspelmeyer2014} M. Aspelmeyer, T. J. Kippenberg, and F. Marquardt, “Cavity optomechanics,”
\textit{Rev. Mod. Phys.} \textbf{86}, 1391 (2014).

\bibitem{Wang2014} G. Wang, L. Huang, Y.-C. Lai, and C. Grebogi, “Nonlinear dynamics
and quantum entanglement in optomechanical systems,” \textit{Phys. Rev. Lett.}
\textbf{112}, 110406 (2014).

\bibitem{Corbitt2007} T. Corbitt, Y. Chen, E. Innerhofer, H. Müller-Ebhardt, D. Ottaway, H. Rehbein,
D. Sigg, S. Whitcomb, C. Wipf, and N. Mavalvala, “An all-optical
trap for a gram-scale mirror,” \textit{Phys. Rev. Lett.} \textbf{98}, 150802 (2007).

\bibitem{Metzger2004} C. H. Metzger and K. Karrai, “Cavity cooling of a microlever,” \textit{Nature}
\textbf{432}, 1002 (2004).

\bibitem{Kleckner2006} D. Kleckner and D. Bouwmeester, “Sub-kelvin optical cooling of a
micromechanical resonator,” \textit{Nature} \textbf{444}, 75 (2006).

\bibitem{Carmon2005} T. Carmon, H. Rokhsari, L. Yang, T. J. Kippenberg, and K. J. Vahala,
“Temporal behavior of radiation-pressure-induced vibrations of an optical
microcavity phonon mode,” \textit{Phys. Rev. Lett.} \textbf{94}, 223902 (2005).

\bibitem{Thompson2008} J. D. Thompson, B. Zwickl, A. Jayich, F. Marquardt, S. Girvin, and
J. Harris, “Strong dispersive coupling of a high-finesse cavity to a micromechanical
membrane,” \textit{Nature} \textbf{452}, 72 (2008).

\bibitem{Eichenfield2009} M. Eichenfield, J. Chan, R. M. Camacho, K. J. Vahala, and O. Painter,
“Optomechanical crystals,” \textit{Nature} \textbf{462}, 78 (2009).

\bibitem{Aspelmeyer2012} M. Aspelmeyer, P. Meystre, and K. Schwab, “Quantum optomechanics,”
\textit{Phys. Today} \textbf{65}, 29–35 (2012).

\bibitem{van2008} P. van Loock, N. Lütkenhaus, W. Munro, and K. Nemoto, “Quantum
repeaters using coherent-state communication,” \textit{Phys. Rev. A} \textbf{78}, 062319
(2008).

\bibitem{Wang2012} T.-J. Wang, S.-Y. Song, and G. L. Long, “Quantum repeater based
on spatial entanglement of photons and quantum-dot spins in optical
microcavities,” \textit{Phys. Rev. A} \textbf{85}, 062311 (2012).

\bibitem{Ladd2006} T. D. Ladd, P. van Loock, K. Nemoto, W. J. Munro, and Y. Yamamoto,
“Hybrid quantum repeater based on dispersive CQED interactions between
matter qubits and bright coherent light,” \textit{New J. Phys.} \textbf{8}, 184
(2006).

\bibitem{Zhou2011} L. Zhou, Y. Han, J. Jing, and W. Zhang, “Entanglement of nanomechanical
oscillators and two-mode fields induced by atomic coherence,” \textit{Phys.
Rev. A} \textbf{83}, 052117 (2011).

\bibitem{Xu1996} W. Xu, “Emission of acoustic and optical phonons by hot electrons in a
two-dimensional electron system in parallel magnetic fields,” \textit{Phys. Rev.
B} \textbf{54}, 2775 (1996).

\bibitem{Nishioka2014} H. Nishioka, “Stokes suppression and supercontinuum generation
by differential two-phonon excitation,” \textit{Opt. Express} \textbf{22}, 26457–26461
(2014).

\bibitem{James2007} D. James and J. Jerke, “Effective hamiltonian theory and its applications
in quantum information,” \textit{Can. J. Phys.} \textbf{85}, 625–632 (2007).

\bibitem{Gamel2010} O. Gamel and D. F. James, “Time-averaged quantum dynamics and
the validity of the effective hamiltonian model,” \textit{Phys. Rev. A} \textbf{82}, 052106
(2010).

\bibitem{Ralph2001} T. Ralph, A. White, W. Munro, and G. Milburn, “Simple scheme for
efficient linear optics quantum gates,” \textit{Phys. Rev. A} \textbf{65}, 012314 (2001).

\bibitem{Knill2001} E. Knill, R. Laflamme, and G. J. Milburn, “A scheme for efficient quantum
computation with linear optics,” \textit{Nature} \textbf{409}, 46 (2001).

\bibitem{Bergou2005} J. A. Bergou and M. Hillery, “Universal programmable quantum state
discriminator that is optimal for unambiguously distinguishing between
unknown states,” \textit{Phys. Rev. Lett.} \textbf{94}, 160501 (2005).

\bibitem{Scheel2006} S. Scheel, W. J. Munro, J. Eisert, K. Nemoto, and P. Kok, “Feed-forward
and its role in conditional linear optical quantum dynamics,” \textit{Phys. Rev.
A} \textbf{73}, 034301 (2006).
\end{thebibliography}

\end{document}